\documentclass[a4paper,12pt]{article}
\author{Pawel Sikorski}
\usepackage[english]{babel}
\usepackage[utf8]{inputenc}
\usepackage{graphicx} 
\usepackage{graphics}
\usepackage{setspace} 
\usepackage{color}
\usepackage[font={small}]{caption}
\usepackage[hidelinks]{hyperref}
\hypersetup{
    colorlinks,
    linkcolor={red!50!black},
    citecolor={blue!60!black},
    urlcolor={blue!80!black}
}
\usepackage{siunitx}
\usepackage[version=3]{mhchem} 
\usepackage[a4paper,top=2.0cm, bottom=2.5cm, left=2.0cm, right=2.cm]{geometry}

\usepackage{tikz}
\usetikzlibrary{calc,decorations.pathmorphing,patterns,arrows,colorbrewer,arrows.meta,backgrounds}

\usepackage[style=chem-acs,backend=biber,maxnames=10,doi=true,url=true]{biblatex}
\usepackage{datetime}
\usepackage{pgfgantt}

\usepackage{wrapfig}
\DeclareFieldFormat{doi}{\href{https://doi.org/#1}{#1}}
\DeclareBibliographyCategory{needsurl}

\renewbibmacro*{url+urldate}{%
  \ifcategory{needsurl}{
    \href{LINK}{\printfield{url}}%
    \iffieldundef{urlyear}
      {}
      {
       }
       }
    {}}

\usepackage{float}
\usepackage{amsmath}
\usepackage{amssymb}
\usepackage{enumitem}
\usepackage{fancyhdr}
\usepackage[outercaption]{sidecap}
\usepackage{csquotes}
\usepackage{siunitx}
\usepackage{mathrsfs}
\usepackage{fancyvrb}

\usepackage{bm}
\newcommand{\rvec}{\bm{r}}
\newcommand{\kvec}{\bm{k}}

\newcommand{\ee}{\mathrm{e}}
\newcommand{\ii}{\mathrm{i}}

\usepackage{physics}
\usetikzlibrary{arrows.meta,fadings}
\usepackage{pgfplots}
\pgfplotsset{compat=1.8}

\DeclareSIUnit{\angstrom}{\text{\normalfont\AA}}

\providecommand{\Svec}{\bm{S}}
\providecommand{\Shat}{\hat{\bm{S}}}  

\begin{document}
\title{FDTBX -- Computational Tools for Simulation of X-ray Fiber Diffraction Patterns from Atomic Coordinates}
\date{}
\maketitle
\begin{center}
Department of Physics, Norwegian University of Science and Technology (NTNU), Trondheim, Norway
\end{center}

\begin{abstract}
\noindent
Fiber diffraction is one of the few experimental techniques capable of
resolving molecular structure in partially ordered, non-crystalline
systems such as fibrous proteins, biopolymers and synthetic polymers.
Because fiber diffraction patterns are complex, exhibiting
paracrystalline order, packing defects and broad, overlapping
reflections, their interpretation relies on model-based refinement, in
which theoretical patterns simulated from candidate atomic models are
iteratively compared with experiment. We present \texttt{fdtbx}, a
modern, open and extensible Python toolbox for simulating X-ray fiber
diffraction patterns directly from atomic coordinates. Built on the
crystallographic library \texttt{cctbx} and designed for readability
and multicore execution, \texttt{fdtbx} computes structure factors from
a PDB model and constructs realistic reciprocal-space reflection
profiles that incorporate the principal physical broadening mechanisms
of fiber diffraction: finite crystallite size, orientational disorder
(with Gaussian, Lorentzian and Voigt angular peak shapes) and
paracrystalline (second-kind) lattice disorder in Hosemann's
formulation. A per-reflection shell-quadrature scheme evaluates the
required convolutions at arbitrary query points rather than on a fixed
grid, giving direct control over the trade-off between accuracy and
speed, and an analytic Ewald-projection routine maps each sampled
reflection, with an appropriate Lorentz correction, onto a flat
detector to produce a simulated pattern for direct comparison with
measured images. We illustrate the toolbox on $\alpha$-chitin,
cellulose~I$\alpha$ and I$\beta$, cellulose triacetate and A-form DNA,
and outline a practical simulation workflow. By providing a transparent,
well-documented and parallelizable implementation of the specialized
algorithms of fiber diffraction, \texttt{fdtbx} lowers the barrier to
reproducible model-based analysis and serves as both a research and a
teaching resource.
\end{abstract}
\newpage\clearpage
\section{Introduction}

Fiber diffraction is a specialized X-ray diffraction technique that has historically enabled landmark discoveries in structural biology, polymer science and materials science. Unlike single-crystal diffraction, fiber diffraction deals with partially ordered assemblies of molecules arranged in paracrystalline form. The method has been crucial for determining the structures of biologically and industrially significant macromolecules, including the DNA double helix~\cite{Watson1953}; protein secondary structures such as $\alpha$-helices and $\beta$-sheets~\cite{paulingPleatedSheetNew1951, paulingStructureProteinsTwo1951}; fibrous biomaterials such as cellulose~\cite{Sikorski2004,Sikorski2009}, collagen and keratin; synthetic polymers such as polyethylene and polyamides; and even viruses~\cite{stubbsHelicalViruses2012}. The technique provides information on helical symmetry, molecular packing and conformational parameters, which are essential for understanding physical properties, biological function and material performance.

Fiber diffraction patterns are inherently complex because of paracrystalline order, packing defects, broad, weak and overlapping reflections, and limited resolution. These factors make direct interpretation of experimental data challenging. The typical analysis workflow therefore involves model-based refinement, in which theoretical diffraction patterns are simulated from candidate molecular models and iteratively compared with experimental data. This requires specialized algorithms that handle disorder and complex sample texture, and that allow the geometry of the simulated experiment to be verified. Candidate models are typically proposed by combining chemical information (the chemical structure of the polymer) with information from fiber diffraction, which can indicate the helical symmetry of the polymer chain and its crystallographic packing.


Although the scattering theory of fiber diffraction has been extensively developed~\cite{cochranStructureSyntheticPolypeptides1952,fraserConformationFibrousProteins1973,guinierXrayDiffractionCrystals1994,hosemannDirectAnalysisDiffraction1962,stroudDiffractionPolycrystallineFibers1996,franklinSplittingLayerLines1955,klugOrderdisorderTransitionsStructures1958,worthingtonHelicalDiffractionParacrystalline1989}, few general tools exist for calculating fiber diffraction patterns from atomistic models. Existing approaches often rely on specialized code of narrow scope that does not allow detailed investigation of paracrystalline disorder, crystallite-size effects or orientational disorder.
Alenaizan \emph{et al.}~\cite{alenaizanXrayFiberDiffraction2021} calculated diffraction patterns from atomic coordinates with purpose-written Python code. At each pixel of a flat 2D detector, the program calculated the intensity and phase of the X-rays scattered by the heavy atoms of the molecules examined. Cylindrical averaging was approximated by summing patterns calculated for molecules rotated about the fiber axis, and orientational disorder was modeled by averaging simulations of molecules tilted away from the fiber axis. This approach could model scattering from isolated molecules, but not the effects of paracrystalline packing of aligned molecules.
Prodanovic \emph{et al.}~\cite{prodanovicXrayDiffractionNonuniformly2016} extended classical helical diffraction theory to helices whose pitch varies along their length. The Fourier transform, expressed in cylindrical coordinates as a sum over Bessel orders, was evaluated directly from an explicit real-space description of the deformed structure. In the resulting patterns, nonuniform strain shifted, broadened and weakened the reflections, an effect that grows with layer-line order.
Nishiyama \emph{et al.}~\cite{nishiyamaDiffractionNonperiodicModels2012} calculated powder and fiber diffraction patterns directly from finite, all-atom models of cellulose crystallites, without assuming a repeating unit cell. Powder patterns were computed with the Debye scattering equation, summed over all atom pairs, which gives the orientationally averaged intensity directly. Fiber patterns were computed in two steps. First, the intensity scattered by all atoms was calculated on a three-dimensional reciprocal-space grid and cylindrically averaged about the chain axis. Second, the result was convolved with a Gaussian orientation distribution of crystallites about the fiber axis. Because no periodicity is imposed, the effects of finite crystallite size, twisting and surface disorder appear in the patterns automatically.
HELIX, a helical diffraction simulation program developed by the CCP13 collaboration~\cite{knuppHELIXHelicalDiffraction2004}, builds a simple helical model from a few parameters and calculates its fiber diffraction pattern. HELIX displays the model as a radial projection and calculates the pattern as the cylindrically averaged Fourier transform of the subunit coordinates. The program does not account for the curvature of the Ewald sphere, so the effects of fiber tilt cannot be reproduced. Nor does it include lattice sampling from crystalline packing: only patterns from isolated molecules can be calculated.
Makin \emph{et al.}~\cite{Makin2007} calculated diffraction patterns using a ray-tracing approach. The model coordinates are first used to calculate the positions and intensities of all reciprocal-lattice points of an ideal crystal. The intensity of each point is then corrected for crystallite size, rotation about the fiber axis, angular disorientation and diffraction geometry, and traced to the detector plane. This approach produces realistic diffraction patterns, but it is not readily extended to include paracrystalline disorder.

Here we describe \texttt{fdtbx}, a Python toolbox designed for readability, extensibility and parallel execution on multicore machines. It simulates diffraction patterns from atomic coordinate models and incorporates the disorder effects that are important in the analysis of fiber diffraction data.

\section{Theoretical background}
\label{sec:theory}

This section gives a short overview of the Fourier-transform-based diffraction theory implemented in \texttt{fdtbx}, which allows realistic, high-quality simulation of fiber diffraction experiments. The description below, and the implementation, are largely based on classic textbooks in the field, including those by Vainshtein, Guinier, Fraser and MacRae and Hosemann and Bagchi, as well as other sources ~\cite{vainshtein1966diffraction,fraserConformationFibrousProteins1973,guinierXrayDiffractionCrystals1994,hosemannDirectAnalysisDiffraction1962,kasaiXRayDiffractionMacromolecules2005,millaneDiffractionFibres2006}.

\subsection{Scattering vector and scattering amplitude}

Throughout, we use the crystallographic convention, in which wave vectors carry no factor of $2\pi$: the incident and scattered wave vectors $\kvec_{\text{in}}$ and $\kvec_{\text{out}}$ have magnitude $1/\lambda$, and the scattering vector is $\Svec = \kvec_{\text{out}}-\kvec_{\text{in}}$, with
magnitude $\lvert\Svec\rvert = 2\sin\theta/\lambda = 1/d$, where $2\theta$ is the scattering angle.
Reciprocal-lattice vectors satisfy $\bm{a}_i^{*}\cdot\bm{a}_j = \delta_{ij}$. We describe the specimen in a frame whose $z$ axis is the fiber axis $\hat{\bm{u}}$, and use the cylindrical reciprocal coordinates
\begin{equation}
  R = \bigl(S_x^{2}+S_y^{2}\bigr)^{1/2}, \qquad Z = S_z, \qquad \psi = \operatorname{atan2}(S_y, S_x),
  \label{eq:cyl_coords}
\end{equation}
so that $R$ and $Z$ are reciprocal lengths, as is standard in the fiber diffraction literature~\cite{Fraser1976,vainshtein1966diffraction,fraserConformationFibrousProteins1973}.
X-rays are scattered by electrons, so a specimen is described by its continuous electron density $\rho_e(\rvec)$. In the kinematic approximation the scattered amplitude is the three-dimensional Fourier transform of that density,
\begin{equation}
  A(\Svec) = \int_V \rho_e(\rvec)\,\ee^{2\pi\ii\Svec\cdot\rvec}\,\dd^3r
           = \mathcal{F}\!\left[\rho_e\right](\Svec),
  \label{eq:amplitude_FT}
\end{equation}
and the measured quantity is the intensity,
\begin{equation}
  I(\Svec) = \lvert A(\Svec)\rvert^{2} = A(\Svec)\,A^{*}(\Svec).
  \label{eq:intensity_A2}
\end{equation}

In an ideal infinite crystal, the electron density is built by repeating a unit cell on a lattice. The total density is then a sum of cell contributions centered on the lattice sites $\bm{R} = m\bm{a} + n\bm{b} + p\bm{c}$
($m,n,p\in\mathbb{Z}$),
\begin{equation}
  \rho_e(\rvec) = \sum_{\bm{R}} \rho_{\text{cell}}(\rvec - \bm{R}).
  \label{eq:rho_lattice}
\end{equation}
Substituting \eqref{eq:rho_lattice} into \eqref{eq:amplitude_FT}, the amplitude factorizes into two terms: the {lattice sum} $\mathcal{L}(\Svec) = \sum_{\bm{R}}\ee^{2\pi\ii\Svec\cdot\bm{R}}$, which depends only on the crystal geometry, and the {structure factor} $F(\Svec)$, which encodes the contents of a single cell:
\begin{align}
  A(\Svec)
    &= \sum_{\bm{R}} \ee^{2\pi\ii\Svec\cdot\bm{R}}
       \underbrace{\int_{\text{cell}}
         \rho_{\text{cell}}(\bm{u})\,
         \ee^{2\pi\ii\Svec\cdot\bm{u}}\,\dd^3u}_{\displaystyle F(\Svec)}  = \mathcal{L}(\Svec)\,  F(\Svec).
  \label{eq:A_factored}
\end{align}
For a one-dimensional row of $N$ cells with spacing $a$, the lattice sum is a geometric series whose squared modulus is the Laue interference
function,
\begin{equation}
  \lvert\mathcal{L}_{1D}(S_x)\rvert^{2}
    = \left\lvert\sum_{n=0}^{N-1}\ee^{2\pi\ii S_x n a}\right\rvert^{2}
    = \frac{\sin^{2}(\pi N S_x a)}{\sin^{2}(\pi S_x a)} .
  \label{eq:laue1d}
\end{equation}
This function has principal maxima of height $N^{2}$ at $S_x a = h$ ($h\in\mathbb{Z}$) with width $\Delta S \sim 1/(Na)$: the peak width is
the reciprocal of the total row length. As $N\to\infty$ the maxima approach a Dirac comb,
\begin{equation}
  \lvert\mathcal{L}_{1D}(S_x)\rvert^{2}
    \;\xrightarrow{\,N\to\infty\,}\;
    \frac{N}{a}\sum_{h=-\infty}^{\infty}
    \delta\!\left(S_x - \frac{h}{a}\right).
  \label{eq:laue_limit}
\end{equation}
In three dimensions the three Laue conditions must hold simultaneously,
\begin{equation}
  \Svec\cdot\bm{a} = h,\qquad
  \Svec\cdot\bm{b} = k,\qquad
  \Svec\cdot\bm{c} = \ell,\qquad h,k,\ell\in\mathbb{Z},
  \label{eq:laue3d}
\end{equation}
that is, $\Svec$ must coincide with a reciprocal-lattice vector $\bm{D}_{hk\ell} = h\bm{a}^{*} + k\bm{b}^{*} + \ell\bm{c}^{*}$, where
\begin{equation}
   \bm{a}^{*} = \frac{\bm{b}\times\bm{c}}{V_c},\quad
   \bm{b}^{*} = \frac{\bm{c}\times\bm{a}}{V_c},\quad
   \bm{c}^{*} = \frac{\bm{a}\times\bm{b}}{V_c},\quad
   V_c = \bm{a}\cdot(\bm{b}\times\bm{c}).
   \label{eq:recip_basis}
\end{equation}

The structure factor is
\begin{equation}
  F_{hk\ell} = \sum_{j=1}^{s} f_j(\bm{D}_{hk\ell})\,
               \ee^{2\pi\ii(hx_j + ky_j + \ell z_j)},
  \label{eq:structure_factor}
\end{equation}
where $f_j$ is the atomic scattering factor and $(x_j,y_j,z_j)$ are the fractional coordinates of atom $j$.
The Bragg intensity is $I_{hk\ell}\propto\lvert F_{hk\ell}\rvert^{2}$. In \texttt{fdtbx}, $|F_{hk\ell}|$ is calculated with the crystallographic library \texttt{cctbx} \cite{grosse-kunstleveComputationalCrystallographyToolbox2002}. Since
$\lvert\bm{D}_{hk\ell}\rvert = 1/d_{hk\ell}$ and $\lvert\Svec\rvert = 2\sin\theta/\lambda$, the Laue condition reduces to
Bragg's law,
\begin{equation}
  \frac{2\sin\theta}{\lambda} = \frac{1}{d_{hk\ell}}
  \quad\Longrightarrow\quad
  2\,d_{hk\ell}\sin\theta = \lambda .
  \label{eq:bragg}
\end{equation}

\subsection{Finite crystallite size}
\begin{figure}[htbp]
\centering
\includegraphics[width=.5\linewidth]{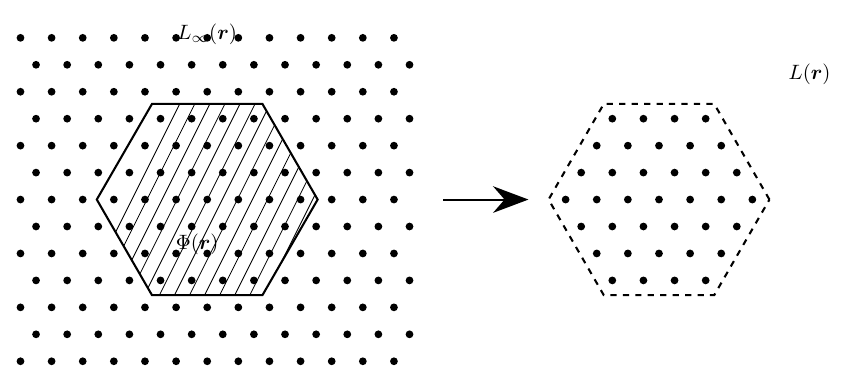}
\caption{Action of the shape function $\Phi(\rvec)$. The infinite lattice $L_\infty(\rvec)$ (left) is multiplied by $\Phi(\rvec)$, which is unity inside the
crystallite region (hexagon) and zero outside, yielding the finite crystallite $L(\rvec)$ (right). In reciprocal space this product becomes
a convolution, smearing each sharp reflection by $\widetilde{\Phi}(\Svec)$ [Eq.~\eqref{eq:shape_convolution}].}
\label{fig:shape}
\end{figure}

Fiber diffraction samples often consist of assemblies of molecules that form small crystalline regions. The Fourier-transform approach shows how this affects the diffraction pattern~\cite{vainshtein1966diffraction}. A finite crystallite is obtained by multiplying the infinite lattice $L_\infty(\rvec)$\footnote{Referred to as $A(\rvec)$ by Vainshtein~\cite{vainshtein1966diffraction}.} by a {shape function} $\Phi(\rvec)$ that is unity inside the crystallite and zero outside, so that $L(\rvec) = L_\infty(\rvec)\,\Phi(\rvec)$
(Fig.~\ref{fig:shape}). By the convolution theorem, the amplitude becomes the ideal infinite-lattice transform $\mathcal{L}_\infty(\Svec)$ {convolved} with the transform of the shape function,
  \begin{equation}
    A(\Svec) = \big[F(\Svec)\,\mathcal{L}_\infty(\Svec)\big]
              * \widetilde{\Phi}(\Svec),
    \qquad \widetilde{\Phi} = \mathcal{F}[\Phi].
    \label{eq:shape_convolution}
  \end{equation}
In one dimension this is the finite geometric series already summed in \eqref{eq:laue1d}. More generally, each infinitely sharp reciprocal-lattice point is smeared by $\widetilde{\Phi}(\Svec)$\footnote{Referred to as $S(\Svec)$ by Vainshtein~\cite{vainshtein1966diffraction}.}, which broadens every reflection to a width of order $1/L$ for a crystallite of linear size $L$. This size broadening is {independent} of the order of the reflection, a feature that distinguishes it from the paracrystalline disorder described below. Near a single Bragg point the sharply peaked lattice sum is well approximated by a $\operatorname{sinc}$-type profile (see \autoref{sec:shell_quadrature}).

The finite-crystal amplitude $A(\Svec)$ is exact, but the measured quantity is the intensity $I(\Svec) = \lvert A(\Svec)\rvert^{2}$. Because $\lvert G * \widetilde{\Phi}\rvert^{2} \ne
\lvert G\rvert^{2} * \lvert\widetilde{\Phi}\rvert^{2}$ for a general amplitude $G = F\,\mathcal{L}_\infty$, the intensity cannot be written exactly as a convolution. 
\begin{equation}
  I(\Svec)
    \;\approx\;
    \big[\,\lvert F(\Svec)\rvert^{2}\,
           \lvert\mathcal{L}_\infty(\Svec)\rvert^{2}\,\big]
    * \lvert\widetilde{\Phi}(\Svec)\rvert^{2}.
  \label{eq:intensity_conv}
\end{equation}

\subsection{Paracrystalline disorder}
\label{sec:paracrystal}

Fiber specimens are very often composed of small, disordered crystals, whose reflections broaden with increasing scattering angle in a way that finite size alone cannot explain. Following the fiber diffraction literature~\cite{vainshtein1966diffraction,fraserConformationFibrousProteins1973,hosemannDirectAnalysisDiffraction1962}, it is useful to distinguish two kinds of lattice disorder.

\emph{Disorder of the first kind} leaves the lattice itself intact: every scatterer is displaced independently from a fixed ideal site by a
random amount with a fixed variance (thermal motion, static point disorder). It multiplies the reflections by a Debye--Waller factor (DW), in anisotropic form
\begin{equation}
DW(\Svec) = \exp{-\frac{1}{4}\sum\limits_{i=1}^3 B_i  (\Svec \cdot \hat{\bm{a}}_i)^2}
\end{equation}
or in isotropic form
\begin{equation}
DW(\Svec) = \exp{-2\pi^2 \left< u^2\right> S^2},
\end{equation}
where $\left< u^2\right>$ is the isotropic mean-square atomic displacement along the scattering vector, in \AA$^2$, and $B = 8\pi^2 \left< u^2\right>$.
This factor attenuates high-angle intensity but leaves the sharp Bragg peaks, and their positions and widths, unchanged; the lost coherent intensity is transferred into a diffuse background.

\emph{Disorder of the second kind}, the {paracrystal}, destroys long-range order itself. The vector between neighboring
lattice points is a random variable, and because displacements accumulate along the lattice, the positional uncertainty of a site grows
{broaden progressively}, with a width that grows roughly as the square of the reflection order,
\begin{equation}
  \Delta S_{h} \;\sim\; \pi^{2} g^{2} h^{2}\,\frac{1}{a},
  \label{eq:paracrystal_width}
\end{equation}
where the paracrystallinity parameter $g = \sqrt{\langle\Delta a^2\rangle}/a$ is the relative standard deviation of the nearest-neighbor spacing. High-order reflections eventually broaden into the background and vanish, which places an intrinsic resolution limit on paracrystalline fibers.

Lattice disorder of the second kind (paracrystalline disorder, in Hosemann's sense) does not attenuate or blur the structure factor or the shape function; it changes the lattice itself, and so enters the amplitude
\begin{equation}
    A(\bm{S}) = \bigl[\,F(\bm{S})\,\mathcal{L}_\infty(\bm{S})\,\bigr]
                * \widetilde{\Phi}(\bm{S}),
    \qquad \widetilde{\Phi} = \mathcal{F}[\Phi],
    \label{eq:shape_convolution_para}
\end{equation}
as a modification of the ideal lattice factor $\mathcal{L}_\infty$. In a paracrystal each cell is placed relative to its neighbour with a random error, so positional errors accumulate and no ideal reference lattice exists. For a one-dimensional lattice along the unit-cell vector $\bm{a}_i$, Hosemann's ensemble-averaged lattice factor is  
\begin{equation}
    Z_{i,\mathrm{para}}(\bm{S})
    = \frac{1-|\phi_i|^2}
           {1-2\,|\phi_i|\cos\!\bigl(2\pi\,\bm{a}_i\!\cdot\!\bm{S}\bigr)+|\phi_i|^2},
    \label{eq:Z}
\end{equation}
where $\phi_i$ is the Fourier transform of the nearest-neighbour spacing distribution. The full three-dimensional replacement is the product over the three axes, 

\begin{equation}
|\mathcal{L}_\infty(\bm{S})|^2\to  Z_{i,\mathrm{para}}(\bm{S})=\prod_{i=1}^{3} Z_{i,\mathrm{para}}(\bm{S}).
\end{equation}
And the sample-averaged intensity $\left<I(\Svec)\right>$  becomes:

\begin{equation}
 \left<I(\Svec)\right> \approx \left[|F_\text{hkl}|^2 \;  Z_{i,\mathrm{para}}(\bm{S})\right] * |\widetilde{\Phi}(\bm{S})|^2
\end{equation}
For a Gaussian spacing distribution with mean spacing $d_i$ and standard deviation $\sigma_{d,i}$, the relative fluctuation $g_i \equiv \sigma_{d,i}/d_i$ (Hosemann's $g$-factor) gives
\begin{equation}
|\phi_i(\bm{S})| = \exp[-2\pi^2 g_i^2\,(\bm{a}_i\!\cdot\!\bm{S})^2] 
\end{equation}
as $g_i\to 0$ the factor $Z_i$ collapses to the ideal lattice of sharp peaks.

At the reflection of integer order $(h,k,l)$, with $h_i$ the component along axis
$i$, the periodicity of the cosine in \eqref{eq:Z} means the entire local peak is
governed by the value of the coherence factor at the reflection,
\begin{equation}
    A_i \equiv |\phi_i(h_i)| = \exp\bigl(-2\pi^2 g_i^2\,h_i^2\bigr),
    \qquad 0 < A_i \le 1
    \label{eq:Ah}
\end{equation}
whose $h_i^2$ dependence is the signature of second-kind disorder: the effect
grows with reflection order. Two exact properties follow from \eqref{eq:Z}. The
peak height at the reflection is $Z_i(h_i)=(1+A_i)/(1-A_i)$, and, since the
integral of $Z_i$ over one period in $u_i\equiv\bm{a}_i\!\cdot\!\bm{S}$ is unity,
the integral breadth is
\begin{equation}
    \beta_i = \frac{1}{Z_i(h_i)} = \frac{1-A_i}{1+A_i}
    \qquad\text{(in units of } u_i\text{).}
    \label{eq:breadth}
\end{equation}
As the order rises, $A_i$ falls, the peak height drops and the breadth grows.

\begin{figure}[htbp]
\centering
\includegraphics[width=.7\linewidth]{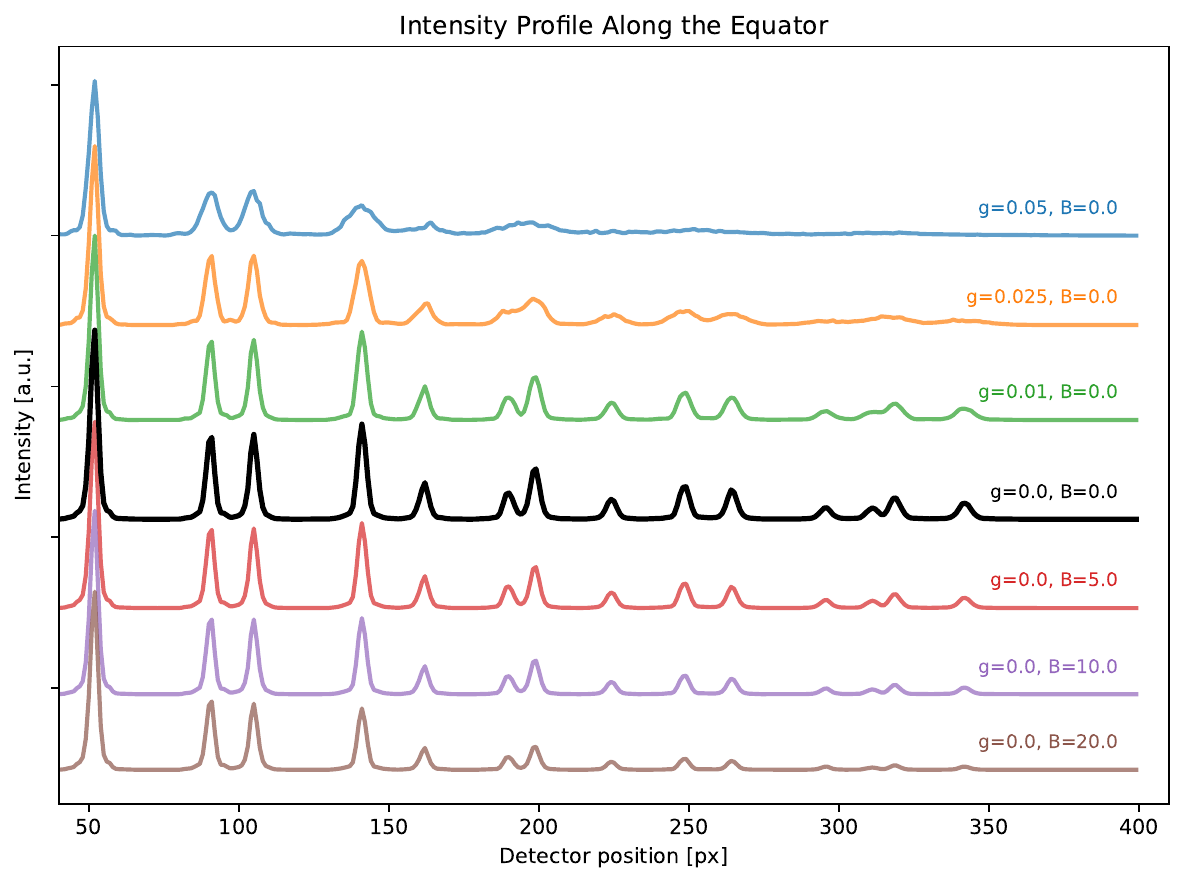}
\caption{Equatorial intensity profile for various disorder parameters, assuming isotropic $B$ and $g$. Model unit cell with $a = b = \SI{10}{\angstrom}$, $c = \SI{5}{\angstrom}$, $\gamma = \SI{120}{\degree}$ and $|F_{hk\ell}|=1$.}
\label{fig:equator}
\end{figure}

The three broadening mechanisms can be distinguished by how the width of a reflection scales with its order $h$: finite size gives a width that is {constant} in $h$; disorder of the first kind gives {no} broadening but a Debye--Waller attenuation; and disorder of the second kind gives a width that grows as $h^{2}$. The effects of the latter two are shown in \autoref{fig:equator}, where equatorial intensity profiles simulated with $|F_{hk\ell}|=1$ are plotted for several values of the disorder parameters $B$ and $g$. Real fiber data combine all three broadening mechanisms. \texttt{fdtbx} allows them to be disentangled quantitatively by forward simulation from a model, which, to our knowledge, existing fiber diffraction simulation software does not support.

\subsection{Orientational disorder: convolution on the sphere}
\label{sec:disorientation}

In fiber diffraction, fiber texture describes a preferred orientation in which a particular crystallographic direction of each crystallite (or molecule) is parallel to the fiber axis, while the crystallites are randomly rotated about that shared axis. Because of this random azimuthal rotation, the diffraction pattern is equivalent to that obtained by rotating a single crystal about the axis. Real fibers are not perfectly aligned: the texture axes of the individual crystallites are scattered by small angles about the mean fiber direction. This {disorientation} smears each reflection along an arc. Mathematically, it is a convolution in reciprocal space between the already-broadened Bragg peak and an orientation distribution.

For a single reflection, combining the structure factor with the finite-size shape function gives a localized peak profile near the reciprocal-lattice point $\bm{D}_{hk\ell}$,
\begin{equation}
  p(\Svec - \bm{D}_{hk\ell})
    = \big\lvert F(\bm{D}_{hk\ell})\big\rvert^{2}\,
      \big\lvert\widetilde{\Phi}(\Svec - \bm{D}_{hk\ell})\big\rvert^{2},
  \label{eq:single_peak}
\end{equation}
centered on $\bm{D}_{hk\ell}$ with a width $\sim 1/L$ set by the crystallite size $L$.
Tilting a crystallite by a rotation $\mathcal{R}$ rotates its entire reciprocal lattice rigidly, so the peak at $\bm{D}$ moves to
$\mathcal{R}\bm{D}$, carrying its internal profile $\widetilde{\Phi}$ unchanged. A fiber is an incoherent ensemble of crystallites whose orientations are distributed according to a normalized function $g(\mathcal{R})$, so the measured intensity is the ensemble average
\begin{equation}
  I(\Svec) = \int p\big(\Svec - \mathcal{R}\bm{D}\big)\,
             g(\mathcal{R})\,\dd\mathcal{R},
  \label{eq:orientation_average}
\end{equation}
with the integral running over the rotation group with its invariant measure. Equation~\eqref{eq:orientation_average} is already a convolution, but over rotations rather than over $\Svec$; the remaining step is to turn it into a convolution in reciprocal space.

A reflection lies at radius $D = \lvert\bm{D}\rvert = 1/d_{hk\ell}$ and at polar angle $\Theta$ from the fiber ($z$) axis, with axial and radial coordinates (Eq.~\eqref{eq:cyl_coords})
\begin{equation}
  Z = D\cos\Theta, \qquad R = D\sin\Theta .
  \label{eq:polar}
\end{equation}
Disorientation of the fiber axis spreads the crystallite $c$-axes over a small polar angle $\beta$ with an azimuthally symmetric distribution
$N(\beta)$, normalized by $\int N(\beta)\,\dd\Omega = 1$. Because a rigid rotation preserves $\lvert\bm{D}\rvert$, the peak is displaced
{along the arc of constant radius} $D$: disorientation moves intensity in the angular coordinate $\Theta$ while leaving $D$ fixed
(a Debye--Scherrer-type arc). Provided the peak is narrow compared with the orientation spread (the usual situation in fibers, $1/L \ll D\,\Delta\beta$), the profile is effectively rigid along the arc, and the ensemble average \eqref{eq:orientation_average} reduces to a convolution over the arc angle,
\begin{equation}
  I(D,\Theta) \;=\; \int p\big(D,\,\Theta-\beta\big)\,
                    N(\beta)\,\dd\beta
             \;=\; \big(p * N\big)(D,\Theta).
  \label{eq:arc_convolution}
\end{equation}
Each reflection is thus the size-broadened peak {convolved with the orientation distribution} $N$, spread into an arc whose angular width is that of $N$. The convolution theorem gives a direct practical consequence: angular variances add,
\begin{equation}
  \sigma_{\Theta}^{2} \;=\; \sigma_{\text{size}}^{2}
                          + \sigma_{\text{orient}}^{2},
  \label{eq:variance_add}
\end{equation}
where $\sigma_{\text{size}}\approx 1/(LD)$ is the angular width subtended by the size-broadened peak at radius $D$. A measured arc width is therefore the size broadening and the disorientation added in quadrature. For a Gaussian model,
$N(\beta)\propto\exp(-\beta^{2}/2\sigma_{\text{orient}}^{2})$, the arc is Gaussian; other choices (e.g.\ a Lorentzian) enter \eqref{eq:arc_convolution} unchanged.

\subsection{Angular peak shapes}
Crystallite orientation is described by a distribution of directions about the direction of the reciprocal-lattice vector $\bm{D}_{hk\ell}$, whose unit vector we write as $\Shat = \bm{D}_{hk\ell}/\lvert\bm{D}_{hk\ell}\rvert$.
Writing $\Delta$ for the angle between a sampled direction $\hat{\bm{n}}$ and $\Shat$, the angular distribution enters the shell convolution as a weight $g(\Delta)$. Three shapes are implemented. Unlike $N(\beta)$ above, which is normalized to unit integral, they are written here normalized to unit peak height, $g(0)=1$, which is convenient for comparing shapes; the normalization to unit integral that \eqref{eq:orientation_average} requires is applied in the quadrature weights (\autoref{sec:shell_quadrature}). In the code, $\sigma_\theta$ and $\gamma_\theta$ are set by the parameters {\tt sigma\_sphere} and {\tt gamma\_sphere}.

The Gaussian shape models a mosaic spread with short tails,
\begin{equation}
    g_{\mathrm{G}}(\Delta) = \exp\!\left(-\frac{\Delta^{2}}{2\sigma_\theta^{2}}\right),
\end{equation}
where $\sigma_\theta$ is the angular standard deviation. Its full width at half
maximum is
\begin{equation}
    \mathrm{FWHM}_{\mathrm{G}} = 2\sqrt{2\ln 2}\,\sigma_\theta
    \approx 2.3548\,\sigma_\theta .
\end{equation}

Because $g$ describes a distribution of {directions}, it is a two-dimensional density over the sphere, and the profile observed along an arc is its one-dimensional cut (marginal) through the reflection. The Lorentzian and Voigt shapes are therefore defined as two-dimensional distributions whose marginal along an arc is the familiar one-dimensional profile. 
Gaussian and Lorentzian line shapes and their convolution, the Voigt function, are the standard description of peak profiles in diffraction~\cite{langfordRapidMethodAnalysing1978,youngProfileShapeFunctions1982,thompsonRietveldRefinementDebyeScherrer1987}, and Gaussian orientation distributions are the usual model of disorientation in fiber diffraction~\cite{holmesEffectDisorientationIntensity1974,stubbsEffectDisorientationIntensity1974,millaneDiffractionFibres2006}. Because $g$ describes a distribution of directions, the Lorentzian and Voigt shapes are defined here as two-dimensional distributions on the sphere whose profile along an arc is the familiar one-dimensional function. The Lorentzian shape, which models a long-tailed spread, is
\begin{equation}
    g_{\mathrm{L}}(\Delta) = \left(1+\frac{\Delta^{2}}{\gamma_\theta^{2}}\right)^{-3/2},
\end{equation}
which gives an arc with half width at half maximum $\gamma_\theta$, so that $\mathrm{FWHM}_{\mathrm{L}} = 2\gamma_\theta$. The one-dimensional Lorentzian $\gamma_\theta^{2}/(\Delta^{2}+\gamma_\theta^{2})$ cannot be used directly, because its integral over solid angle diverges. The Voigt shape $g_{\mathrm{V}}$ is the two-dimensional convolution of the Gaussian and Lorentzian components, normalized to $g_{\mathrm{V}}(0)=1$ and evaluated numerically. Its profile along an arc is the one-dimensional Voigt function, and it reduces to $g_{\mathrm{G}}$ for $\gamma_\theta\to0$ and to $g_{\mathrm{L}}$ for $\sigma_\theta\to0$. An orientation distribution with a Gaussian core (mosaic spread) and Lorentzian wings is therefore described by the single pair $(\sigma_\theta,\gamma_\theta)$.

\subsection{Numerical evaluation: per-reflection shell quadrature}
\label{sec:shell_quadrature}

\begin{figure}[htbp]
\centering
\includegraphics[width=.3\linewidth, trim=3cm 2cm 3cm 2cm, clip]{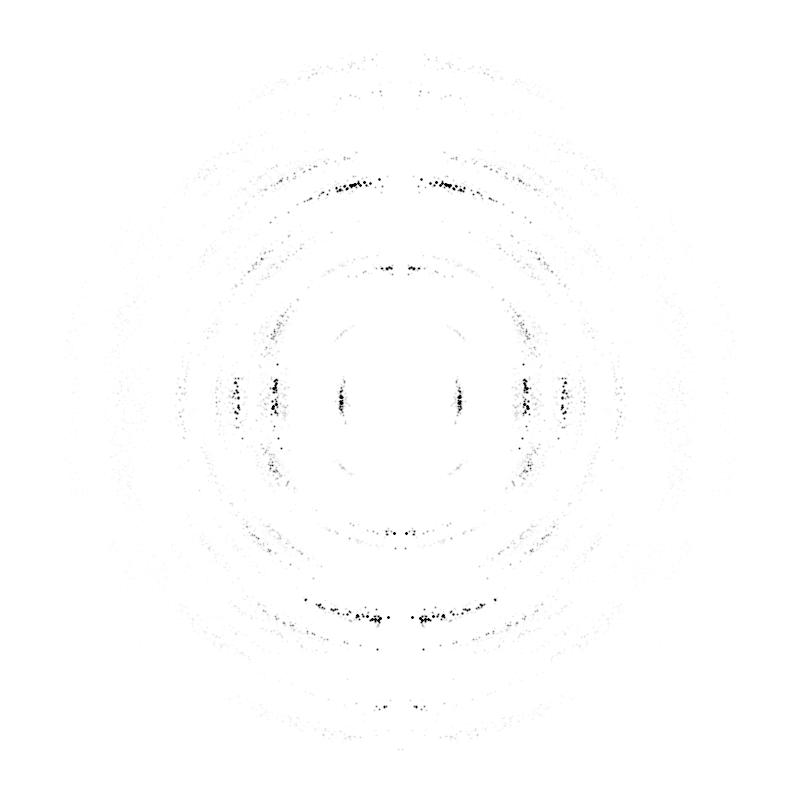}
\includegraphics[width=.3\linewidth, trim=3cm 2cm 3cm 2cm, clip]{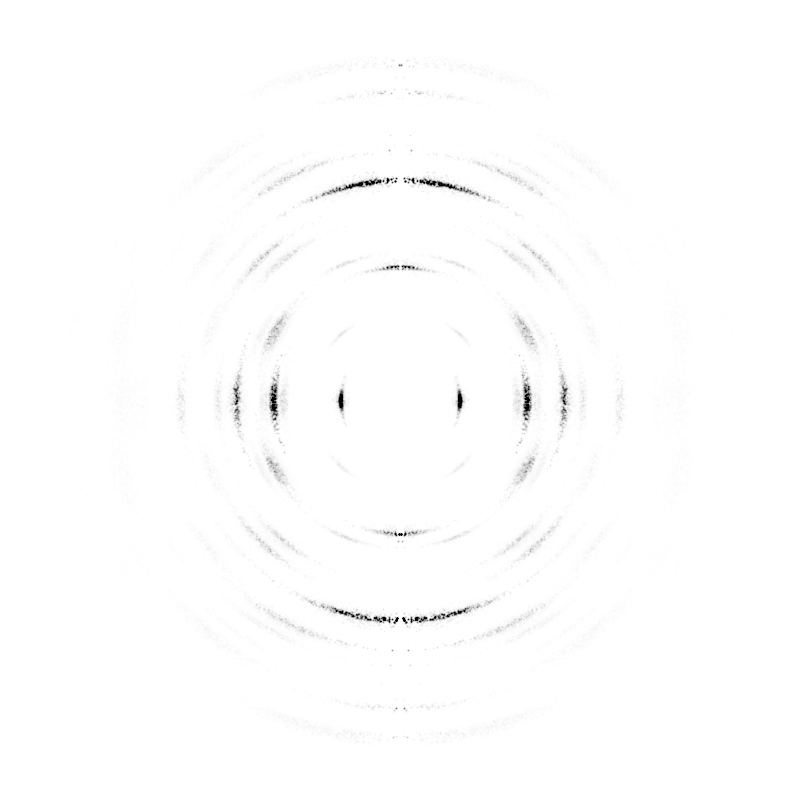}\\
\includegraphics[width=.3\linewidth, trim=3cm 2cm 3cm 2cm, clip]{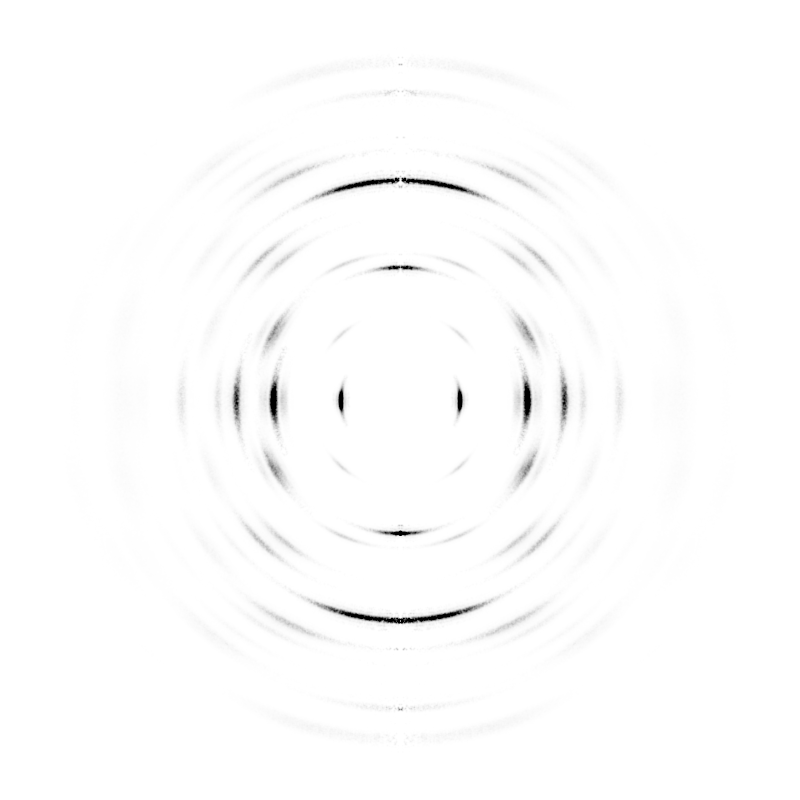}
\includegraphics[width=.3\linewidth, trim=3cm 2cm 3cm 2cm, clip]{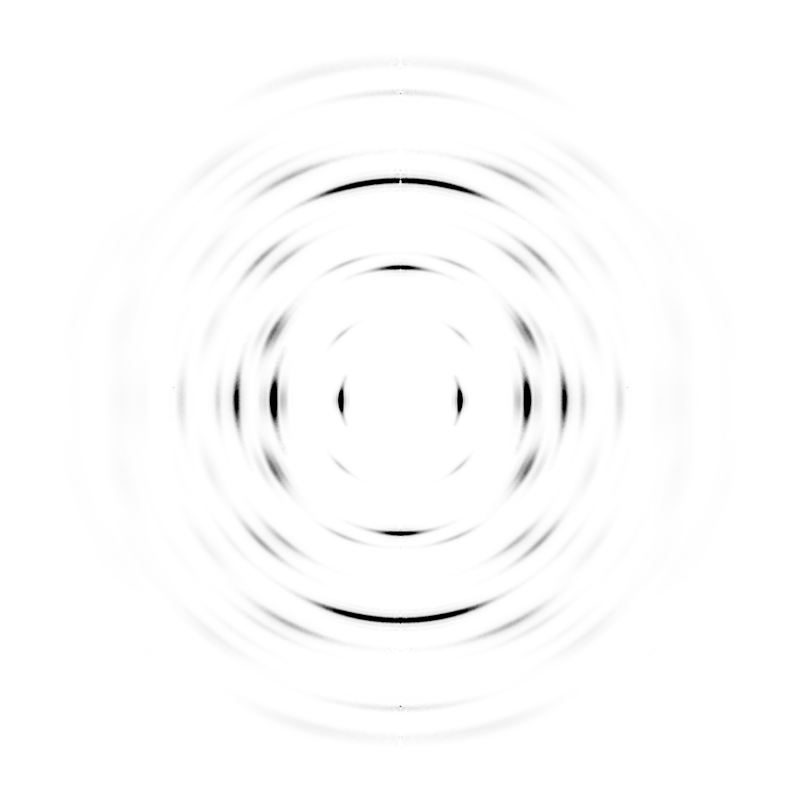}
\caption{Fiber diffraction patterns of cellulose triacetate (CTA) (unit cell: $a=\SI{5.939}{\angstrom}$, $b=\SI{11.431}{\angstrom}$, $c=\SI{10.46}{\angstrom}$, $\alpha=\SI{90}{\degree}$, $\beta=\SI{90}{\degree}$, $\gamma=\SI{95.4}{\degree}$)~\cite{Sikorski2004} calculated with (left to right) 50, 500, 5000 and 50\,000 sampling points per $hk\ell$ reflection. At low sampling densities each reflection appears as a scatter of discrete points on the detector; as the number of points increases, the projected intensity converges to continuous reflection profiles.} 
\label{fig:CTA}
\end{figure}

The convolution \eqref{eq:arc_convolution} must be evaluated at the reciprocal-space points that a given experiment actually samples. This subsection summarizes the numerical scheme used in \texttt{fdtbx}, which treats each reflection independently and evaluates the convolution directly at arbitrary query points rather than on a regular grid. This gives direct control over the trade-off between accuracy and speed. \autoref{fig:CTA} shows the same pattern calculated for different numbers of query points.
We call this approach {shell quadrature}: the convolution of the orientation distribution, a density on the sphere $\lvert\bm{s}\rvert = \lvert\bm{D}_{hk\ell}\rvert$, with the Laue kernel is evaluated as a weighted sum of kernels centred on quadrature nodes placed on that sphere, with weights proportional to the orientation probability and the solid angle each node represents. Numerical quadrature refers to calculating the approximate value of a definite integral using weighted sums. 

Around an isolated reflection at $\bm{D}_{hk\ell}$, the finite-crystallite Laue shape function is the product of three one-dimensional
factors,
\begin{equation}
 K(\Svec)
    = \big\lvert\widetilde{\Phi}(\Svec)\big\rvert^{2}
    = \prod_{i\in\{a,b,c\}}
        \frac{\sin^{2}\!\big(\pi N_i\,\bm{a}_i\!\cdot\!\Svec\big)}
             {\big(\pi\,\bm{a}_i\!\cdot\!\Svec\big)^{2}}  \frac{1}{N_i},
  \label{eq:I_shape}
\end{equation}
written here in its non-periodic (single-peak) envelope, with peak value $N_aN_bN_c$ and first zeros at $\bm{a}_i\!\cdot\!\Svec = 1/N_i$, i.e.\ at a Cartesian half-width of $1/(N_i\lvert\bm{a}_i\rvert)$ from the center. The orientation distribution is represented as
a {shell}: intensity concentrated on the sphere $\lvert\Svec\rvert = r_0 = \lvert\bm{D}_{hk\ell}\rvert$ and distributed in angle about $\Shat=\bm{D}_{hk\ell}/r_0$ according to $g$. The observed
reflection is their convolution,
\begin{equation}
  I(\Svec) = \int G_{\text{shell}}(\bm{s})\,
                  K(\Svec - \bm{s})\,\dd^3s .
  \label{eq:shell_conv}
\end{equation}

Two features of the physically relevant regime make \eqref{eq:shell_conv} cheap to evaluate exactly. First, the shell has no radial extent. Because a rigid rotation preserves $\lvert\bm{D}_{hk\ell}\rvert$, orientational disorder has no radial component: the orientation distribution is a surface density on the sphere $\lvert\bm{s}\rvert = r_0$, and the radial profile of a reflection is set entirely by the kernel $K$, which already contains the finite crystallite size through $N_i$. The shell is therefore
\begin{equation}
  G_{\text{shell}}(\bm{s})
  = \frac{\delta\!\left(\lvert\bm{s}\rvert - r_0\right)}{r_0^{2}}\,
    \frac{g(\Delta)}{\displaystyle\int g(\Delta)\,\dd\Omega},
  \label{eq:gshell}
\end{equation}
where $g$ is the angular profile of the previous subsection (Gaussian, Lorentzian or Voigt) and $\Delta$ is the angle from $\Shat$. The factor $r_0^{-2}$ cancels the $r^{2}$ of the volume element, so that $\int G_{\text{shell}}\,\dd^3s = 1$ and the integrated intensity of each reflection is conserved, as required by \eqref{eq:orientation_average}. The volume integral \eqref{eq:shell_conv} thus reduces exactly to a {surface} integral over the sphere, and no radial discretization of the shell is needed.

Second, the kernel is {compact}: truncated at a few zeros per lattice direction, its support is a small parallelepiped, so the integral at any query point draws only on a local patch of the shell.

The sphere is parametrized by tangent-plane offsets $\hat{\bm{n}}(d_1,d_2)$ about $\Shat$, with orthonormal tangent vectors $\bm{e}_1,\bm{e}_2$,
\begin{equation}
  \hat{\bm{n}}(d_1,d_2)
    = \frac{\Shat + d_1\bm{e}_1 + d_2\bm{e}_2}
           {\sqrt{1 + d_1^{2} + d_2^{2}}},
  \qquad \bm{s} = r_0\,\hat{\bm{n}} .
  \label{eq:gnomonic}
\end{equation}
Two distinct sets of points are used for each reflection. The first is a set of quadrature nodes $\bm{s}_k$ on the shell, which serve only to evaluate the integral \eqref{eq:shell_conv}; the second is a set of randomly sampled query points $\Svec_i$, at which the intensity is evaluated and which are projected onto the detector (described below). The quadrature nodes are placed on a fixed, regular $(d_1,d_2)$ grid, so that the integral is evaluated deterministically and its accuracy is controlled by the grid spacing, which is tied to the kernel width. The grid covers the angular patch $\lvert d_{1,2}\rvert \le d_{\max}$, with $d_{\max} = n_{\text{ang}}\sqrt{\sigma_\theta^{2}+\gamma_\theta^{2}}$ ($n_{\text{ang}}=6$ by default). Each node carries the weight
\begin{equation}
  w_k = \frac{g(\Delta_k)\,J_k}{\sum_j g(\Delta_j)\,J_j},
  \qquad
  J = \underbrace{\left(1+d_1^{2}+d_2^{2}\right)^{-3/2}}_{\text{solid-angle Jacobian}},
  \label{eq:node_weight}
\end{equation}
where $\Delta = \arccos\!\big[(1+d_1^{2}+d_2^{2})^{-1/2}\big]$ is the exact great-circle angle from $\Shat$. The normalization ensures $\sum_k w_k = 1$ for every reflection, independently of $r_0$, of the angular widths $\sigma_\theta$ and $\gamma_\theta$, of the grid spacing and of the patch size; it is the discrete counterpart of the denominator in \eqref{eq:gshell}. For a Gaussian profile the patch contains essentially all of $g$. For a Lorentzian or Voigt profile, whose solid-angle integral grows logarithmically with the patch size, the normalization is taken over the patch, so that $d_{\max}$ defines the extent of the modeled orientation distribution. The nonlinear
normalization in \eqref{eq:gnomonic} and the exact Jacobian in \eqref{eq:node_weight} are retained in full; no small-angle expansion is
made anywhere. 

Because the scheme is a direct quadrature rather than a transform, the convolution can be evaluated at {any} set of Cartesian query points $\{\bm{p}_m\}$,
\begin{equation}
  I(\bm{p}_m) \;\approx\; \sum_{k\,:\,(\bm{p}_m - \bm{s}_k)
                                \in\,\operatorname{supp} K}
                  w_k\, K(\bm{p}_m - \bm{s}_k) ,
  \label{eq:direct_sum}
\end{equation}
where the sum runs only over the nodes $\bm{s}_k$ that lie within the compact support of the kernel around $\bm{p}_m$. The pipeline exploits this property: each
reflection is processed on its own shell, at its own radius $r_0$, and evaluated only where needed. \autoref{fig:CTA} shows that reflection intensity does not depent on the number od query points used in the calculation. 
 
Each reciprocal-space point is handled separately and mapped to the detector by an independent geometric routine that returns the pixel
position corresponding to that $\Svec$. The two operations are cleanly decoupled: \eqref{eq:direct_sum} supplies the {intensity} at a
reciprocal-space point, while the projection routine supplies its {position} on the detector, accounting for the experimental
geometry (tilt, detector distance, orientation and beam parameters). The projection routine also performs the cylindrical averaging, by determining the crystal orientation(s) for which a given reciprocal-space point satisfies the diffraction condition (see \autoref{sec:ewald-analytical}). Because the intensity evaluation imposes no grid, the query points can be chosen to suit the projection. Unlike the quadrature nodes, they are not placed on a grid: they are drawn at random, by Monte Carlo sampling of the reflection's shell in $(r,d_1,d_2)$ coordinates,
\begin{equation}
  \Svec_i = (r_0 + \varrho_i)\,\hat{\bm{n}}(d_{i1}, d_{i2}),
  \qquad
  d_{i1},\, d_{i2} \sim p_d, \quad \varrho_i \sim p_\varrho,
  \qquad i = 1,\dots,M,
  \label{eq:mc_points}
\end{equation}
where $p_\varrho$ is a radial density whose width is set by the kernel and $p_d$ is a tangent-plane density whose width is set by the orientation distribution. The sampling density is not the intensity: the points only have to cover the region in which $I$ is non-negligible. To make the accumulated pattern an unbiased estimate of $I(\Svec)$, independent of the choice of $p_d$, $p_\varrho$ and $M$, each point carries the importance weight
\begin{equation}
  W_i = \frac{1}{M\,p(\Svec_i)}
      = \frac{(r_0+\varrho_i)^{2}\,J(d_{i1},d_{i2})}
             {M\,p_d(d_{i1})\,p_d(d_{i2})\,p_\varrho(\varrho_i)},
  \label{eq:mc_weight}
\end{equation}
where $J$ is the Jacobian of \eqref{eq:node_weight} and the factor $r^{2}J$ converts the density in $(r,d_1,d_2)$ to a density per unit reciprocal-space volume. The intensity $I(\Svec_i)$ at each random point is computed from the fixed quadrature grid through \eqref{eq:direct_sum}, so the randomness enters only through where the points are placed. Then $\sum_i W_i\,f(\Svec_i) \to \int f(\Svec)\,\dd^3S$ for any function $f$ supported within the sampled region; in particular, the integrated intensity of each reflection is recovered, and the profile shape is not distorted by the sampling density. 
Without $W_i$ the accumulated density would be $p(\Svec)\,I(\Svec)$ rather than $I(\Svec)$, which narrows both the arcs and the radial profiles. To keep the weights bounded, the sampling densities are matched to the profiles they sample. Radially, $p_\varrho$ is a Gaussian whose standard deviation equals the half-width of the kernel along $\Shat$, evaluated with the effective, paracrystal-broadened sizes $N_{\text{eff},i}$. Angularly, its standard deviation is $1.3$ times the combined width of the orientation distribution and the tangential half-width of the kernel. Because the kernel is truncated to a parallelepiped $\lvert\bm{a}_i\cdot\Delta\Svec\rvert\le n_{\text{pad}}/N_i$, whose corners can lie several half-widths from the center for oblique cells, each density is taken as a mixture, $(1-\epsilon)\,p_{\text{narrow}} + \epsilon\,p_{\text{broad}}$ with $\epsilon = 0.2$, in which the broad component covers the whole kernel support. This bounds $W_i$ everywhere the intensity is non-zero; without it, rare points in the corners of the support receive very large weights and appear as isolated bright pixels. Each sampled point is then projected individually. The statistical error of the accumulated intensity decreases as $M^{-1/2}$, which is the convergence seen in \autoref{fig:CTA}.

\begin{figure}[htbp]
\centering
\includegraphics[width=.49\linewidth, trim=2cm 2cm 2cm 2cm, clip]{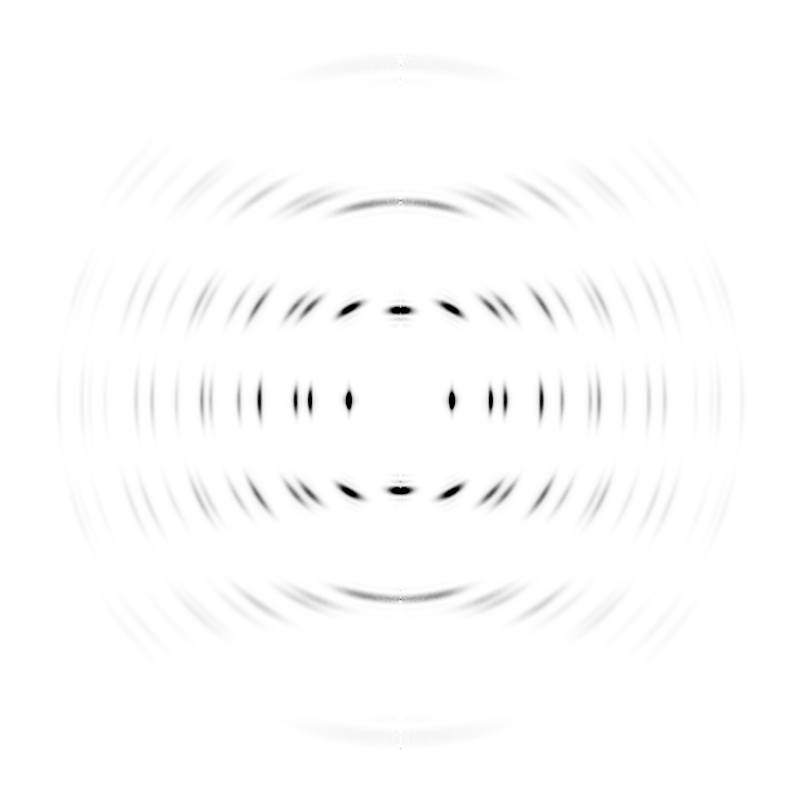}
\includegraphics[width=.49\linewidth, trim=2cm 2cm 2cm 2cm, clip]{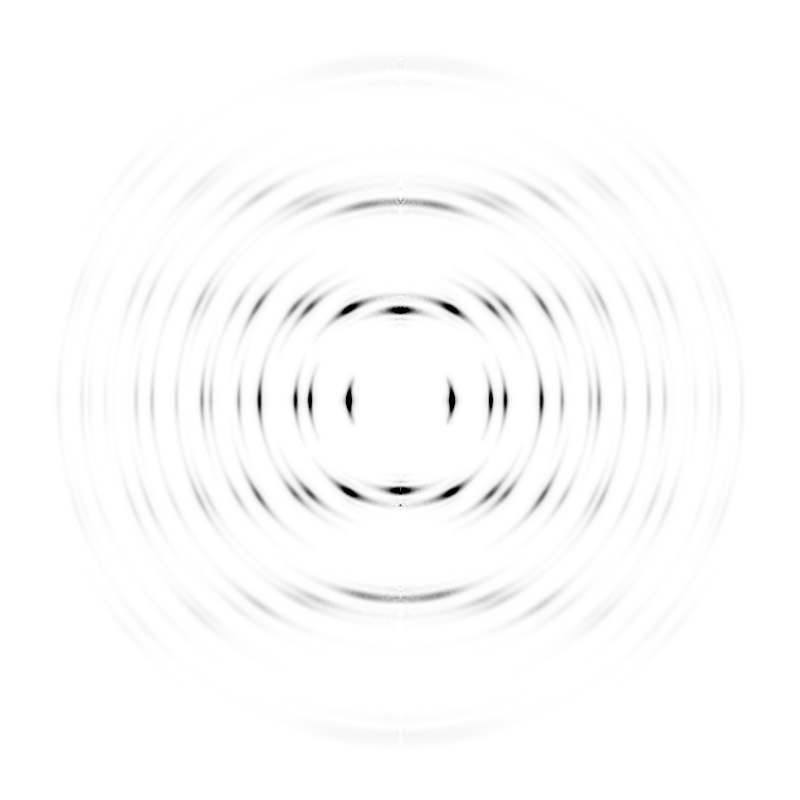}
\caption{Fiber diffraction patterns calculated with a Gaussian (left; {\tt sigma\_sphere}~$=\SI{4}{\degree}$, {\tt gamma\_sphere}~$=\SI{0}{\degree}$) and a Lorentzian (right; {\tt sigma\_sphere}~$=\SI{0}{\degree}$, {\tt gamma\_sphere}~$=\SI{4}{\degree}\sqrt{2 \ln 2}$) crystallite orientation distribution. Model unit cell with $a = b = \SI{10}{\angstrom}$, $c = \SI{5}{\angstrom}$, $\gamma = \SI{120}{\degree}$ and $|F_{hk\ell}|=1$. Setting both {\tt sigma\_sphere} and {\tt gamma\_sphere} greater than \SI{0}{\degree} can be used to generate reflections with a Voigt profile.}
\label{fig:GvL}
\end{figure}


\subsection{Lorentz correction on a sampled reciprocal lattice}

The measured intensity of a reflection is not simply proportional to the squared structure factor: it also carries a purely geometric weighting that accounts for how the diffracting volume element is distributed in reciprocal space and how it intersects the Ewald sphere. This weighting is the Lorentz factor.

In \texttt{fdtbx} the reciprocal lattice is not treated as a set of discrete Bragg points. Instead, each reflection is represented by a cloud of sampling points $\{\Svec_i\}$ generated by the convolution routines, which
distribute intensity over a finite reciprocal-space volume to model peak broadening from finite crystallite size, lattice disorder and disorientation. The Lorentz correction must therefore be evaluated {per sampling point} rather than per reflection, since neighboring points within one broadened peak may have appreciably different geometric weights, most notably in the
immediate vicinity of the meridian.

For a sample with fiber symmetry, crystallites are randomly rotated about a common axis $\hat{\bm{u}}$, the fiber axis. Rotational averaging about $\hat{\bm{u}}$ smears every sampling point $\Svec_i$ into a ring in reciprocal space. Writing $D_i = |\Svec_i|$ for the magnitude of the $i$-th sampling vector, the polar angle
$\Theta_i$ between $\Svec_i$ and the fiber axis (as in \autoref{sec:disorientation}) follows from
\begin{equation}
  \cos\Theta_i = \frac{\Svec_i\cdot\hat{\bm{u}}}{D_i},
  \qquad
  \sin^{2}\Theta_i = 1 - \frac{(\Svec_i\cdot\hat{\bm{u}})^{2}}{D_i^{2}} ,
\end{equation}
and the radius of the corresponding rotation ring is
\begin{equation}
  R_i = D_i\,\sin\Theta_i ,
  \label{eq:radius}
\end{equation}
i.e.\ simply the component of $\Svec_i$ perpendicular to $\hat{\bm{u}}$, the cylindrical coordinate $R$ of Eq.~\eqref{eq:cyl_coords}. The intensity associated with the sampling point is distributed uniformly around a ring of circumference $2\pi R_i$, so the intensity per unit length of ring, and hence the recorded intensity, scales as
\begin{equation}
  L_i \propto \frac{1}{R_i} .
\end{equation}

The $1/R_i$ dependence diverges as $\Theta_i \to 0$, that is, for sampling points approaching the meridian, where the point lies on the fiber axis. On a sampled lattice this divergence is not merely a formal inconvenience: meridional and near-meridional peaks always contain sampling points with arbitrarily small $R_i$, so an unregularized factor would assign them unbounded weight, and the summed intensity of the peak would depend on the sampling density. The correction must therefore be regularized.

In the standard treatment of the rotation method, the Lorentz factor is the time a reciprocal-space point spends in the diffraction condition, which is inversely proportional to the component of its velocity normal to the Ewald sphere~\cite{lipsonTrigonometricIntensityFactors2006,buergerCorrectionXrayDiffraction1940}. For a point rotated about the fiber axis $\hat{\bm{u}}$ into the diffraction condition, at the rotated position $\Svec_{\phi,i}$, this normal component per unit rotation angle is
\begin{equation}
  \bigl\lvert \Svec_{\phi,i}\cdot(\hat{\kvec}_{\text{in}}\times\hat{\bm{u}})\bigr\rvert
  = R_i\,\bigl\lvert\hat{\kvec}_{\text{out},i}\cdot\hat{\bm{\psi}}_i\bigr\rvert ,
\end{equation}
i.e.\ the ring radius $R_i$ times the obliquity with which the ring crosses the sphere ($\hat{\bm{\psi}}_i$ is the ring tangent and $\hat{\kvec}_{\text{out},i}$ the sphere normal). Regularized so that it remains bounded, the Lorentz factor is
\begin{equation}
  L_i = \frac{\delta}{\bigl\lvert \Svec_{\phi,i}\cdot(\hat{\kvec}_{\text{in}}\times\hat{\bm{u}})\bigr\rvert + \delta},
  \label{eq:lorentz}
\end{equation}
which reduces to the familiar $1/R_i$ dependence of fiber diffraction~\cite{Fraser1976,millaneDigitalProcessingXray1986} where rings cross the Ewald sphere at normal incidence. On the meridian of the detector the rings only touch the sphere, $\Svec_{\phi,i}\cdot(\hat{\kvec}_{\text{in}}\times\hat{\bm{u}})\to 0$, and the points stay in the diffraction condition longest; without the obliquity term, the projected intensity of every meridional reflection would fall to zero there. $L_i$ is evaluated separately for each of the two solutions of the diffraction condition (\autoref{sec:ewald-analytical}).

The correction is applied pointwise, together with the importance weight \eqref{eq:mc_weight}, so that each sampling point contributes
\begin{equation}
  I^{\mathrm{corr}}_i = W_i\, L_i \, I(\Svec_i)
  \label{eq:lorentz_apply}
\end{equation}
to the detector, and the observable diffraction pattern is obtained by accumulating these contributions onto the detector grid. Because $L_i$ is a smooth, bounded function of $\Svec_i$, this ordering is
self-consistent: the correction can be applied either before or after the convolution kernel is summed, provided it is evaluated at the sampled positions themselves.

\subsection{Analytical solution of the Ewald diffraction condition}
\label{sec:ewald-analytical}

The calculated $I(\Svec)$ can be mapped onto the detector in two ways. The first is to compute the intensity cylindrically averaged about the fiber axis, $I(R,Z)$, in the coordinates of Eq.~\eqref{eq:cyl_coords}, and then map this intensity onto the detector~\cite{Fraser1976}:
\begin{equation}
  I(R,Z) = \frac{1}{2\pi} \int_{0}^{2\pi} I\!\left( R\cos\psi,\, R\sin\psi,\, Z \right) \dd\psi .
  \label{eq:cyl-average}
\end{equation}
The second, used in \texttt{fdtbx}, is to solve the diffraction condition for each sampling point and project the resulting points onto the detector plane. As in a rotation diffraction experiment, a scattering vector $\Svec$ must be rotated about the fiber axis $\hat{\bm{u}}$ until it meets the Ewald sphere for a signal to appear on the detector. In practice, the sample contains many crystallites with random azimuthal orientations about the fiber axis, and the condition is satisfied for only a subset of them. Elastic scattering requires the scattered wave vector to have the same magnitude as the incident one,
\begin{equation}
  \lvert \kvec_{\text{out}} \rvert
  = \lvert \kvec_{\text{in}} + \Svec_\phi \rvert
  = \frac{1}{\lambda},
  \label{eq:ewald}
\end{equation}
where $\Svec_\phi$ is $\Svec$ rotated by an angle $\phi$ about $\hat{\bm{u}}$. With the convention of \autoref{sec:theory}, Eq.~\eqref{eq:ewald} states that $\Svec_\phi$ ends on the Ewald sphere of radius $1/\lambda$ centred at $-\kvec_{\text{in}}$. The rotation angle $\phi$ is a change of the azimuth $\psi$ of Eq.~\eqref{eq:cyl_coords}; $R$ and $Z$ are unchanged by it.

The task is to find, for each $\Svec$, the rotation angle(s) $\phi$ that satisfy Eq.~\eqref{eq:ewald}. Rather than scanning $\phi$ numerically, \texttt{fdtbx} solves for $\phi$ in closed form, which makes vectorized evaluation over many $\Svec$ efficient; there are two solutions when the ring of radius $R$ crosses the Ewald sphere, one when it touches it, and none otherwise.  Each solution gives a scattered wave vector $\kvec_{\text{out}} = \kvec_{\text{in}} + \Svec_\phi$, which is mapped to a pixel position on a flat detector.

\section{Example simulations}

\begin{figure}[htbp]
\centering
\begin{tikzpicture}
\node[anchor=north] (a) at (0,0)
  {\includegraphics[width=.9\linewidth,trim=0cm 0.0cm 0cm 1.9cm, clip]{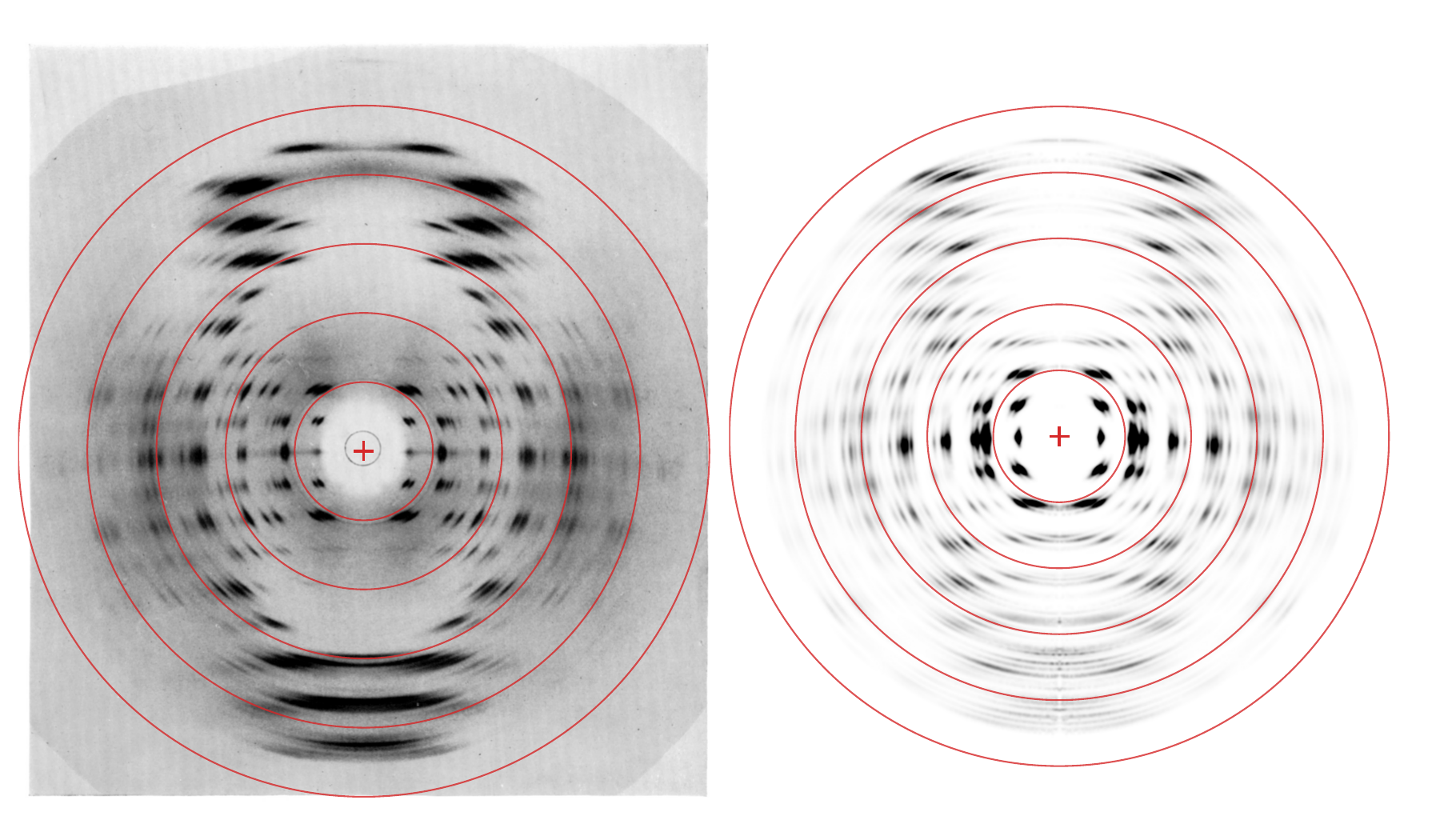}};
\end{tikzpicture}
\caption{
Comparison between the experimental diffraction pattern and the pattern simulated with \texttt{fdtbx} for the sodium salt of A-form DNA. The experimental pattern, from Fuller et al.\ (1965), was recorded at 75\% relative humidity with $\lambda = \SI{1.54}{\angstrom}$~\cite{fullerMolecularConfigurationDeoxyribonucleic1965}.
The simulation used the coordinates reported by Fuller et al.\ (1965), with the structure built using the ``fibre'' tool of the 3DNA package (\url{https://web.x3dna-dssr.org}).
Unit cell: $a=\SI{22.24}{\angstrom}$, $b=\SI{40.62}{\angstrom}$, $c=\SI{28.15}{\angstrom}$, $\alpha=\SI{90}{\degree}$, $\beta=\SI{97}{\degree}$, $\gamma=\SI{90}{\degree}$; space group $C\,1\,2\,1$~\cite{fullerMolecularConfigurationDeoxyribonucleic1965}. The calculation used a tilt angle of \SI{22}{\degree}, equal to the experimental tilt; $g_a=0.01$, $g_b=0.01$, $g_c=0.001$ and $B_a = B_b = B_c = \SI{5.0}{\angstrom\squared}$. Agreement is very good at low angles; the discrepancies on the higher layer lines arise because the coordinate file contains no water molecules, \ce{Na+} or other ions, which contribute high-frequency components to the electron density and therefore affect high-angle intensities most strongly.
Experimental pattern reproduced with permission from Fuller et al., J.\ Mol.\ Biol.\ (1965) 12, 60--80; copyright 1965 Published by Elsevier Ltd.
}
\label{fig:DNA}
\end{figure}

Together, these components define the computational pipeline of the toolbox: build a model assuming an ideal, infinite crystal; calculate the structure factor of each reflection and its shape in reciprocal space, accounting for finite crystallite size, orientational disorder and paracrystalline lattice disorder; and finally project to the detector plane by evaluating the diffraction condition, producing a simulated fiber pattern for comparison with experiment.
Comparisons between simulated and experimental patterns for four semicrystalline polymers ($\alpha$-chitin, cellulose~I$\alpha$, cellulose~I$\beta$ and cellulose triacetate) and for the sodium salt of A-form DNA are shown in \autoref{fig:AChitin} and \autoref{fig:DNA}, respectively. Unit-cell parameters and data sources for each system are given in the figure legends. Notebooks containing the simulation parameters, together with the atomic coordinate files needed to perform the simulations, are distributed with the {\tt fdtbx} code. 

For the simulations shown in \autoref{fig:AChitin} and \autoref{fig:DNA}, the disorder and peak shape parameters, as well as, the global intensity scaling of the calculated patterns were chosen manually to match the experimental and calculate patterns visually. No refinement was carried out. The agreement could be improved further by fitting these parameters to the experimental data.

For the semi-crystalline polymers (\autoref{fig:AChitin}), the agreement between simulated and experimental patterns is very good in the equatorial, off-equatorial and meridional regions of the pattern. For A-form DNA (\autoref{fig:DNA}), the agreement is less satisfactory in the high-angle meridional region, most likely because the atomic coordinate file used in the simulation did not include the water molecules and counter-ions that should be present in the unit cell.

\begin{figure}[htbp]
\centering
\newcommand{\vgap}{-1pt}
\begin{tikzpicture}[img/.style={inner sep=0pt, outer sep=0pt},
                    lbl/.style={anchor=north west, xshift=5pt, yshift=-15pt}]
\node[img, anchor=north] (a) at (0,0)
  {\includegraphics[width=.7\linewidth,trim=0cm 0.2cm 0cm 2.1cm, clip]{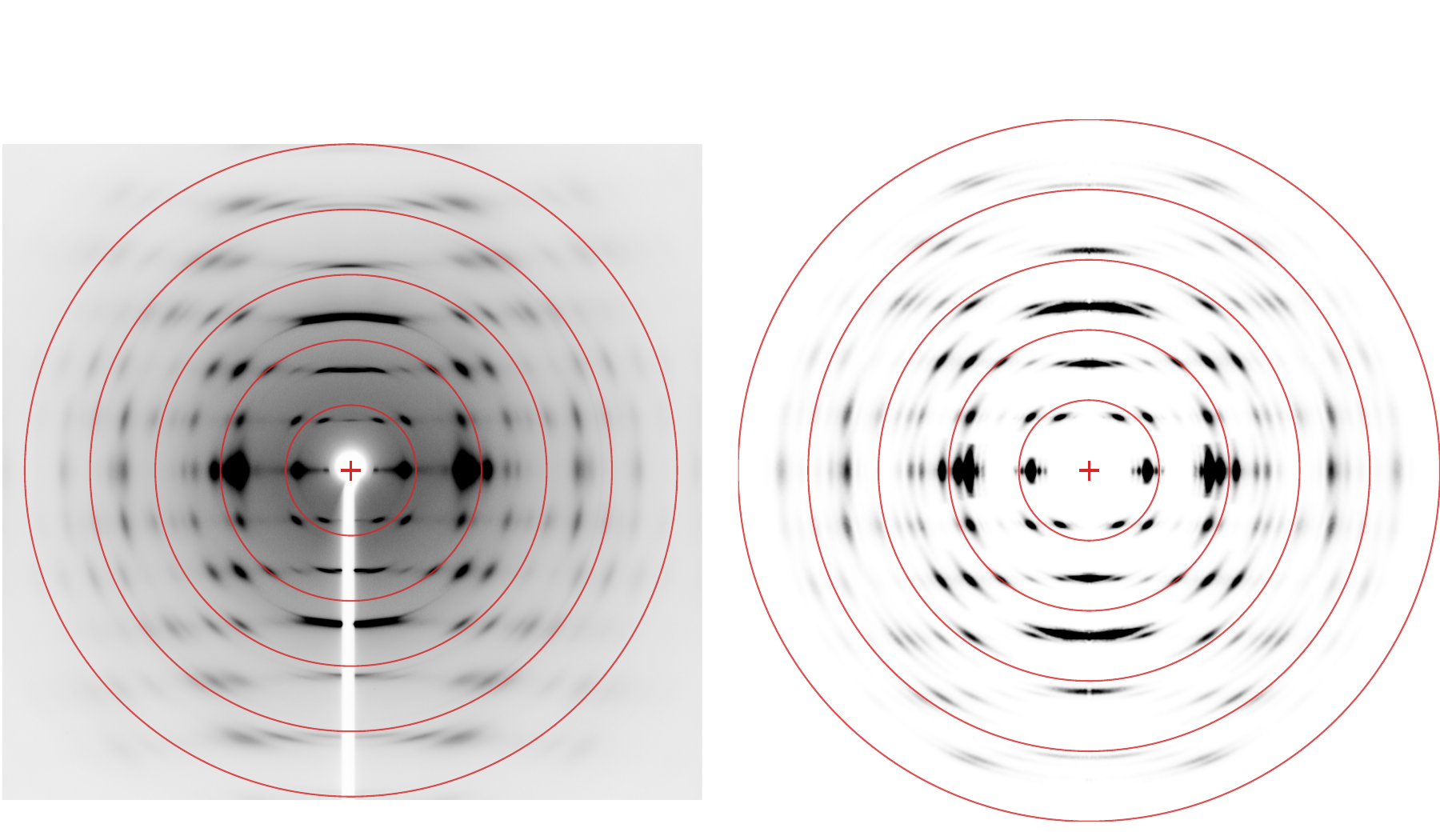}};
\node[img, below=\vgap of a] (b)
  {\includegraphics[width=.7\linewidth,trim=0cm 0.2cm 0cm 2.1cm, clip]{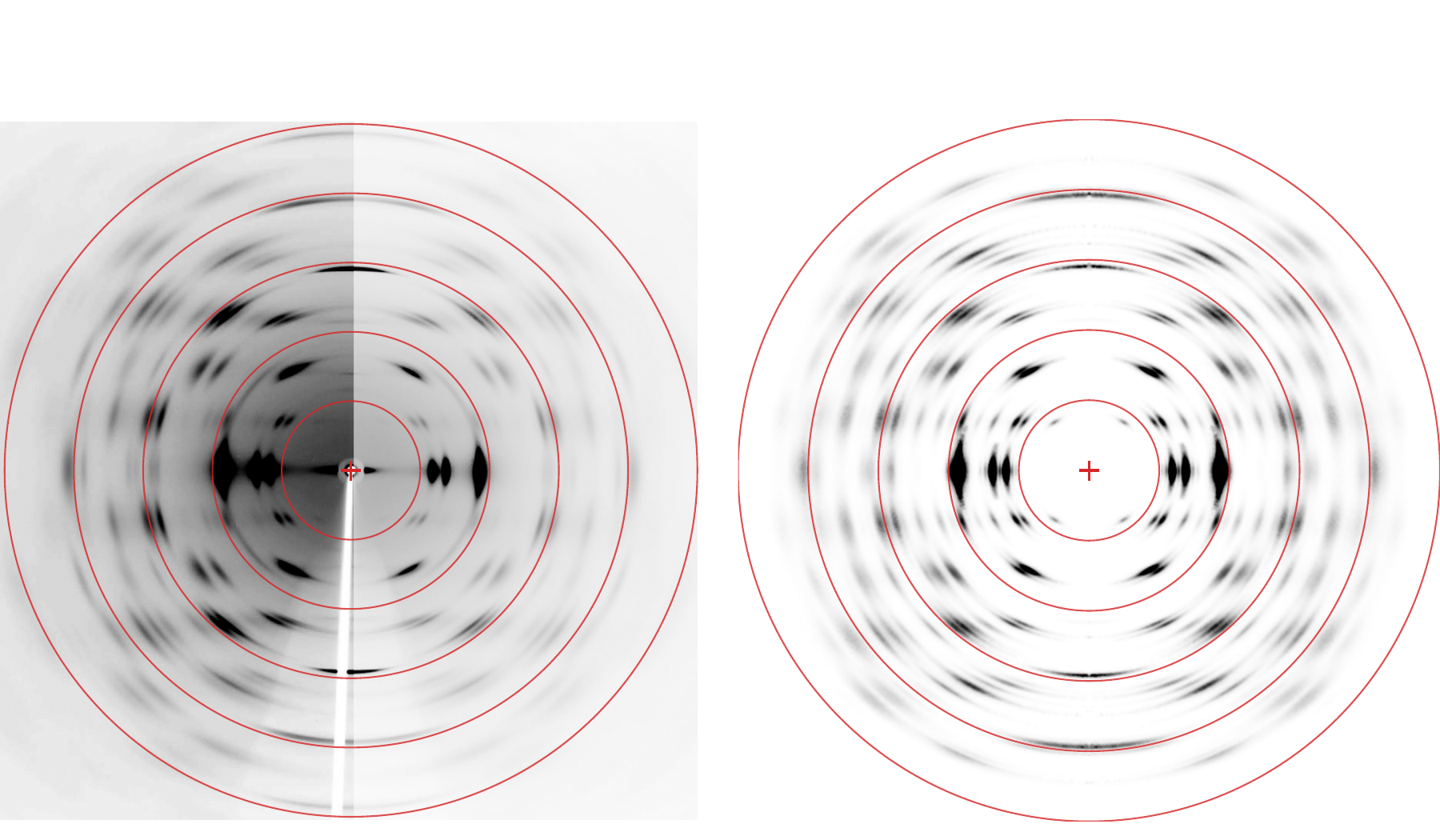}};
\node[img, below=\vgap of b] (c)
  {\includegraphics[width=.7\linewidth,trim=0cm 0.2cm 0cm 2.1cm, clip]{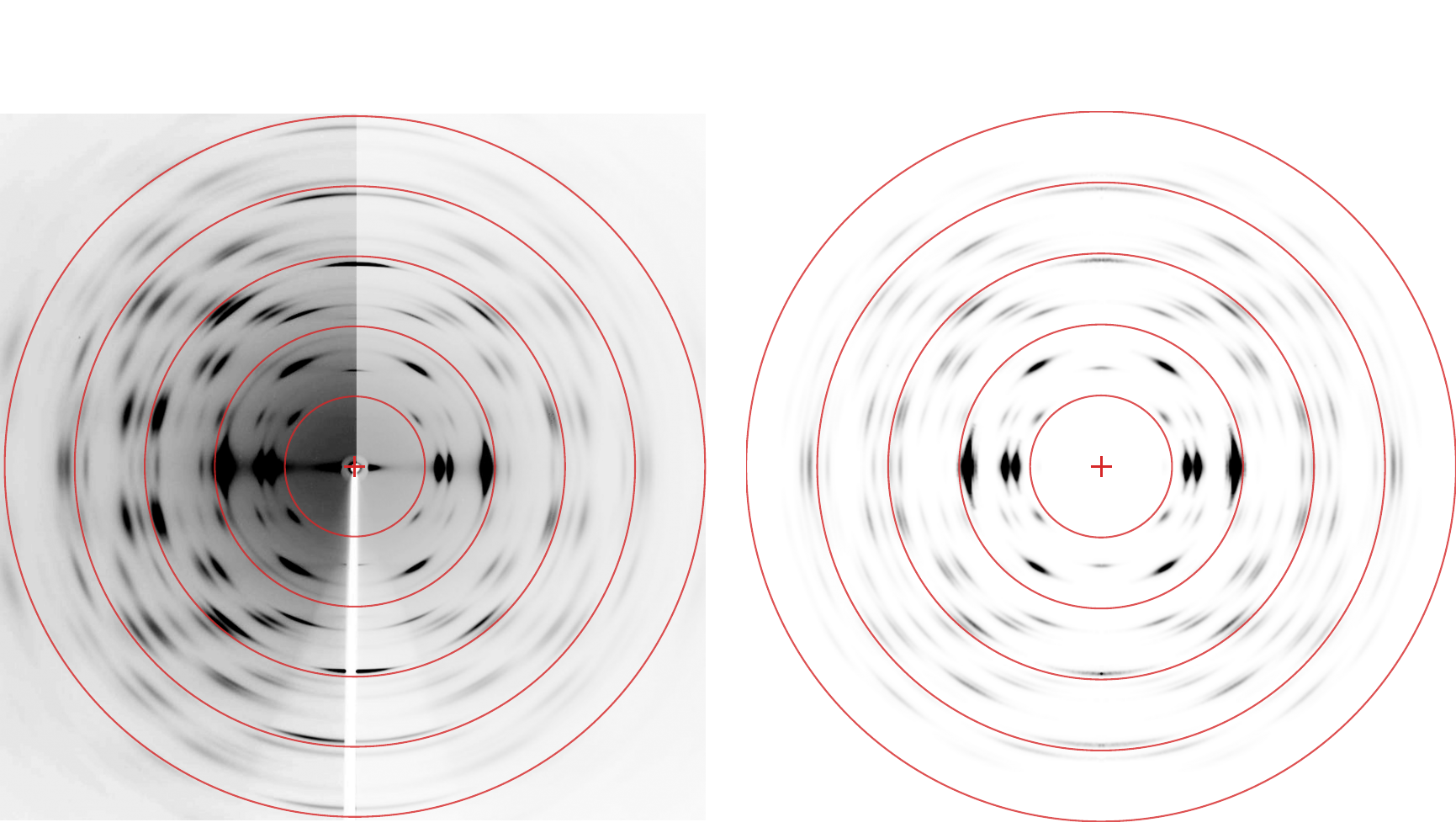}};
\node[img, below=\vgap of c] (d)
  {\includegraphics[width=.7\linewidth,trim=0cm 0.2cm 0cm 2.1cm, clip]{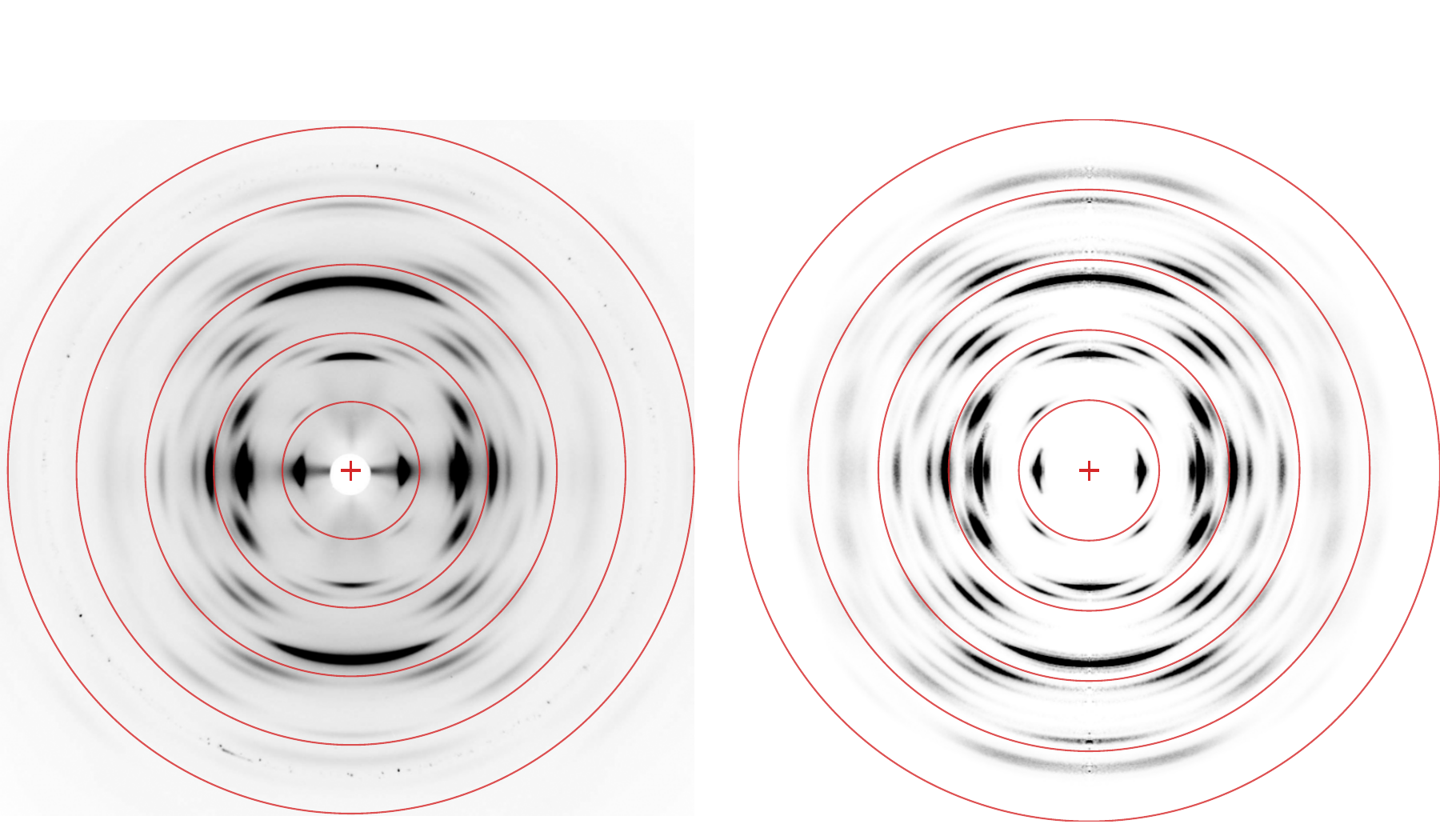}};
\node[lbl] at (a.north west) {(a)};
\node[lbl] at (b.north west) {(b)};
\node[lbl] at (c.north west) {(c)};
\node[lbl] at (d.north west) {(d)};
\end{tikzpicture}
\caption[]{\textit{(continued on next page)}}
\end{figure}

\clearpage
\begin{figure}[H]
  \ContinuedFloat
\caption{Comparison between experimental and calculated fiber diffraction patterns of crystalline polymers. (a) $\alpha$-Chitin, 300\,K structure~\cite{sikorski2009revisit} (unit cell: $a=\SI{4.75}{\angstrom}$, $b=\SI{18.89}{\angstrom}$, $c=\SI{10.333}{\angstrom}$, $\alpha=\SI{90}{\degree}$, $\beta=\SI{90}{\degree}$, $\gamma=\SI{90}{\degree}$); (b) cellulose~I$\alpha$~\cite{nishiyamaCrystalStructureHydrogen2003} (unit cell: $a=\SI{10.4}{\angstrom}$, $b=\SI{6.717}{\angstrom}$, $c=\SI{5.962}{\angstrom}$, $\alpha=\SI{80.37}{\degree}$, $\beta=\SI{118.08}{\degree}$, $\gamma=\SI{114.8}{\degree}$); (c) cellulose~I$\beta$~\cite{nishiyamaCrystalStructureHydrogenBonding2002} (unit cell: $a=\SI{7.784}{\angstrom}$, $b=\SI{8.201}{\angstrom}$, $c=\SI{10.38}{\angstrom}$, $\alpha=\SI{90}{\degree}$, $\beta=\SI{90}{\degree}$, $\gamma=\SI{96.55}{\degree}$); and (d) cellulose triacetate (CTA)~\cite{Sikorski2004} (unit cell: $a=\SI{5.939}{\angstrom}$, $b=\SI{11.431}{\angstrom}$, $c=\SI{10.46}{\angstrom}$, $\alpha=\SI{90}{\degree}$, $\beta=\SI{90}{\degree}$, $\gamma=\SI{95.4}{\degree}$). All calculations were done with the fiber axis normal to the beam (tilt~$=\SI{0}{\degree}$). Experimental pattern in (a) adapted with permission from Sikorski et al., Biomacromolecules 2009, 10, 1100--1105~\cite{sikorski2009revisit}; copyright 2009 American Chemical Society. Experimental data in (b) and (c) kindly provided by Prof.\ Masahisa Wada, Division of Forest and Biomaterials Science, Graduate School of Agriculture, Kyoto University. Experimental pattern in (d) adapted with permission from Sikorski et al., Macromolecules 2004, 37, 4547--4553~\cite{Sikorski2004}; copyright 2004 American Chemical Society.}
\label{fig:AChitin}
\end{figure}
 
\section{Conclusion}
Interpretation of fiber diffraction data can often be a forward-modeling problem: for the many biologically and industrially important assemblies, candidate atomic models can only be assessed by comparing simulated and measured patterns. By providing an implementation of the specialized algorithms this requires, \texttt{fdtbx} lowers the barrier to reproducible model-based experimental data analysis. Building on the Fourier-transform theory of diffraction, \texttt{fdtbx} assembles a pattern from per-reflection structure factors and reciprocal-space shape functions that encode the three broadening mechanisms characteristic of fiber specimens, and then projects each reflection onto the detector.  Future work will extend the library with additional disorder and texture models, and with tools for automated determination of $|F_{hk\ell}|^2$ from experimental fiber diffraction data, that can be accuratelly corrected for texture and disorder effects.

\section*{Code availability}
\texttt{fdtbx} is freely available at \url{https://zenodo.org/records/23189395} under the MIT licence.

\renewcommand*{\bibfont}{\footnotesize}
\begin{spacing}{1}
\begin{footnotesize}

\setlength\bibitemsep{0pt}
\setlength{\biblabelsep}{1pt}
\printbibliography[heading=none]

\end{footnotesize}
\end{spacing}

\end{document}